\documentclass[12pt]{article}

\newcommand{\bl}{\mbox{\boldmath$l$}}
\newcommand{\pbpi}{\,^+\!\mbox{\boldmath$\pi$}}
\newcommand{\mbpi}{\,^-\!\mbox{\boldmath$\pi$}}
\newcommand{\pmbpi}{\,^{\pm}\!\mbox{\boldmath$\pi$}}
\renewcommand{\d}{{\rm d}}

\newcommand{\ppi}{\,^+\!\pi}
\newcommand{\mpi}{\,^-\!\pi}
\newcommand{\pSig}{\,^+\!\Sigma}
\newcommand{\mSig}{\,^-\!\Sigma}
\newcommand{\pmSig}{\,^{\pm}\!\Sigma}
\newcommand{\R}{\bar{R}}
\newcommand{\Om}{\bar{\Omega}}
\newcommand{\pmR}{\,^{\pm}\!R}
\newcommand{\pmbphi}{\,^{\pm}\!\mbox{\boldmath$\phi$}}
\newcommand{\pmbSig}{\,^{\pm}\!\mbox{\boldmath$\Sigma$}}
\newcommand{\pmphi}{\,^{\pm}\!\phi}
\newcommand{\pmbn}{\,^{\pm}\!\mbox{\boldmath$n$}}
\newcommand{\pmbr}{\,^{\pm}\!\mbox{\boldmath$r$}}
\newcommand{\pbv}{\,^{+}\!\mbox{\boldmath$v$}}
\newcommand{\mbv}{\,^{-}\!\mbox{\boldmath$v$}}
\newcommand{\bv}{\mbox{\boldmath$v$}}

\begin{document}

\title{On the possibility of finite quantum Regge calculus}
\author{V.M. Khatsymovsky \\
 {\em Budker Institute of Nuclear Physics} \\ {\em
 Novosibirsk,
 630090,
 Russia}
\\ {\em E-mail address: khatsym@inp.nsk.su}}
\date{}
\maketitle
\begin{abstract}
The arguments were given in a number of our papers that the discrete quantum gravity
based on the Regge calculus possesses nonzero vacuum expectation values of the
triangulation lengths of the order of Plank scale $10^{-33}cm$. These results are
considered paying attention to the form of the path integral measure showing that
probability distribution for these linklengths is concentrated at certain nonzero
finite values of the order of Plank scale. That is, the theory resembles an ordinary
lattice field theory with fixed spacings for which correlators (Green functions) are
finite, UV cut off being defined by lattice spacings. The difference with an ordinary
lattice theory is that now lattice spacings (linklengths) are themselves dynamical
variables, and are concentrated around certain Plank scale values due to {\it
dynamical} reasons.
\end{abstract}

PACS numbers: 04.60.-m Quantum gravity

\newpage

The formal nonrenormalisability of quantum version of general relativity (GR) may
cause us to try to find alternatives to the continuum description of underlying
spacetime structure. An example of such the alternative description may be given by
Regge calculus (RC) suggested in 1961 \cite{Regge}. In RC the exact GR is developed in
the piecewise flat spacetime which is a particular case of general Riemannian
spacetime \cite{Fried}. In turn, the general Riemannian spacetime can be considered as
limiting case of the piecewise flat spacetime \cite{Fein}. Any piecewise flat
spacetime can be represented as collection of a (countable) number of the flat
4-dimensional {\it simplices}(tetrahedrons), and its geometry is completely specified
by the countable number of the freely chosen lengths of all edges (or 1-simplices).
Thus, RC implies a {\it discrete} description {\it alternative} to the usual continuum
one. For a review of RC and alternative discrete gravity approaches see, e. g.,
\cite{RegWil}.

The discrete nature of the Regge's description presents a difficulty in the
(canonical) quantization of such the theory due to the absence of a regular continuous
coordinate playing the role of time. Therefore one cannot immediately develop
Hamiltonian formalism and canonical (Dirac) quantization. To do this we need to return
to the partially continuum description, namely, with respect to only one direction
shrinking sizes of all the simplices along this direction to those infinitely close to
zero. The linklengths and other geometrical quantities become functions of the
continuous coordinate taken along this direction. We can call this coordinate time $t$
and develop quantization procedure with respect to this time. The result of this
procedure can be formulated as some path integral measure. It is quite natural to
consider this measure as a (appropriately defined) limiting continuous time form of a
measure on the set of the original completely discrete Regge spacetimes. This last
completely discrete measure is just the object of interest to be found. The
requirement for this measure to have the known limiting continuous time form can be
considered as a starting postulate in our construction. The issuing principles are of
course not unique, and another approaches to defining quantum measure in RC based on
another physical principles do exist \cite{HamWil1,HamWil2}.

The above condition for the completely discrete measure to possess required continuous
time limit does not defines it uniquely as long as only one fixed direction which
defines $t$ is considered. However, different coordinate directions should be
equivalent and we have a right to require for the measure to result in the canonical
quantization measure in the continuous time limit {\it whatever} coordinate direction
is chosen to define a time. These requirements are on the contrary a priori too
stringent, and it is important that on some configuration superspace (extended in
comparison with superspace of the genuine Regge geometries) such the measure turns out
to exist.

Briefly speaking, we should, first, find continuous time limit for Regge action,
recast it in the canonical Hamiltonian form and write out the Hamiltonian path
integral, the measure in the latter being called for a moment the continuous time
measure; second, we should check for existence and (if exists) find the measure
obeying the property to tend in the continuous time limit (with concept "to tend"
being properly defined) to the found continuous time measure irrespectively of the
choice of the time coordinate direction. When passing to the continuous time RC we are
faced with the difficulty that the description of the infinitely flattened in some
direction simplex purely in terms of the lengths is singular.

The way to avoid singularities in the continuous time limit is to extend the set of
variables via adding the new ones having the sense of angles and considered as
independent variables. Such the variables are the finite rotation matrices which are
the discrete analogs of the connections in the continuum GR. The situation considered
is analogous to that one occurred when recasting the Einstein action in the
Hilbert-Palatini form,
\begin{equation}                                                                    
\label{S-HilPal} {1\over 2}\int{R\sqrt{g}{\rm d}^4x} \Leftarrow {1\over
8}\int{\epsilon_{abcd}\epsilon^{\lambda\mu\nu\rho}e^a_{\lambda}e^b_{\mu}
[\partial_{\nu}+\omega_{\nu},\partial_{\rho}+\omega_{\rho}]^{cd}{\rm d}^4x},
\end{equation}

\noindent where the tetrad $e^a_{\lambda}$ and connection $\omega^{ab}_{\lambda}$ =
$-\omega^{ba}_{\lambda}$ are independent variables, the RHS being reduced to LHS in
terms of $g_{\lambda\mu}$ = $e^a_{\lambda}e_{a\mu}$ if we substitute for
$\omega^{ab}_{\lambda}$ solution of the equations of motion for these variables in
terms of $e^a_{\lambda}$. The Latin indices $a$, $b$, $c$, ... are the vector ones
with respect to the local Euclidean frames which are introduced at each point $x$.

Now in RC the Einstein action in the LHS of (\ref{S-HilPal}) becomes the Regge action,
\begin{equation}                                                                    
\label{S-Regge} \sum_{\sigma^2}{\alpha_{\sigma^2}|\sigma^2|},
\end{equation}

\noindent where $|\sigma^2|$ is the area of a triangle (the 2-simplex) $\sigma^2$,
$\alpha_{\sigma^2}$ is the angle defect on this triangle, and summation run over all
the 2-simplices $\sigma^2$. The discrete analogs of the tetrad and connection, edge
vectors and finite rotation matrices, were first considered in \cite{Fro}. The local
Euclidean frames live in the 4-simplices now, and the analogs of the connection are
defined on the 3-simplices $\sigma^3$ and are the matrices $\Omega_{\sigma^3}$
connecting the frames of the pairs of the 4-simplices $\sigma^4$ sharing the 3-faces
$\sigma^3$. These matrices are the finite SO(4) rotations in the Euclidean case (or
SO(3,1) rotations in the Lorentzian case) in contrast with the continuum connections
$\omega^{ab}_{\lambda}$ which are the elements of the Lee algebra so(4)(so(3,1)) of
this group. This definition includes pointing out the direction in which the
connection $\Omega_{\sigma^3}$ acts (and, correspondingly, the opposite direction, in
which the $\Omega^{-1}_{\sigma^3}$ = $\bar{\Omega}_{\sigma^3}$ acts), that is, the
connections $\Omega$ are defined on the {\it oriented} 3-simplices $\sigma^3$. Instead
of RHS of (\ref{S-HilPal}) we use exact representation which we suggest in our work
\cite{Kha1},
\begin{equation}                                                                    
\label{S-RegCon}%
S(v,\Omega) = \sum_{\sigma^2}{|v_{\sigma^2}|\arcsin{v_{\sigma^2}\circ
R_{\sigma^2}(\Omega)\over |v_{\sigma^2}|}}
\end{equation}

\noindent where we have defined $A\circ B$ = ${1\over 2}A^{ab}B_{ab}$, $|A|$ =
$(A\circ A)^{1/2}$ for the two tensors $A$, $B$; $v_{\sigma^2}$ is the dual bivector
of the triangle $\sigma^2$ in terms of the vectors of its edges $l^a_1$, $l^a_2$,
\begin{equation}                                                                    
\label{v=ll} v_{\sigma^2ab} = {1\over 2}\epsilon_{abcd}l^c_1l^d_2
\end{equation}

\noindent (in some 4-simplex frame containing $\sigma^2$). The curvature matrix
$R_{\sigma^2}$ on the 2-simplex $\sigma^2$ is the path ordered product of the
connections $\Omega^{\pm 1}_{\sigma^3}$ on the 3-simplices $\sigma^3$ sharing
$\sigma^2$ along the contour enclosing $\sigma^2$ once and contained in the
4-simplices sharing $\sigma^2$,
\begin{equation}                                                                    
\label{R-Omega} R_{\sigma^2} = \prod_{\sigma^3\supset\sigma^2}{\Omega^{\pm
1}_{\sigma^3}}.
\end{equation}

\noindent  As we can show, when substituting as $\Omega_{\sigma^3}$ the genuine
rotations connecting the neighbouring local frames as functions of the genuine Regge
lengths into the equations of motion for $\Omega_{\sigma^3}$ with the action
(\ref{S-RegCon}) we get exactly the closure condition for the surface of the 3-simplex
$\sigma^3$ (vanishing the sum of the bivectors of its 2-faces) written in the frame of
one of the 4-simplices containing $\sigma^3$, that is, the identity. This means that
(\ref{S-RegCon}) is the exact representation for (\ref{S-Regge}).

We can pass to the continuous time limit in (\ref{S-RegCon}) in a nonsingular manner
and recast it to the canonical (Hamiltonian) form \cite{Kha2}. This allows us to write
out Hamiltonian path integral. The above problem of finding the measure which results
in the Hamiltonian path integral measure in the continuous time limit whatever
coordinate is chosen as time has solution in 3 dimensions \cite{Kha3}. A specific
feature of the 3D case important for that is commutativity of the dynamical
constraints leading to a simple form of the functional integral. The 3D action looks
like (\ref{S-RegCon}) with area tensors $v_{\sigma^2}$ substituted by the egde vectors
$\bl_{\sigma^1}$ independent of each other. In 4 dimensions, the variables
$v_{\sigma^2}$ are not independent but obey a set of (bilinear) {\it intersection
relations}. For example, tensors of the two triangles $\sigma^2_1$, $\sigma^2_2$
sharing an edge satisfy the relation
\begin{equation}                                                                    
\label{v*v}%
\epsilon_{abcd}v^{ab}_{\sigma^2_1}v^{cd}_{\sigma^2_2} = 0.
\end{equation}

\noindent These purely geometrical relations can be called kinematical constraints.
The idea is to construct quantum measure first for the system with formally
independent area tensors. That is, originally we concentrate on quantization of the
dynamics while kinematical relations of the type (\ref{v*v}) are taken into account at
the second stage. Note that the RC with formally independent (scalar) areas have been
considered in the literature \cite{RegWil,BarRocWil}.

The theory with formally independent area tensors can be called area tensor RC.
Consider the Euclidean case. The Einstein action is not bounded from below, therefore
the Euclidean path integral itself requires careful definition. Our result for the
constructed in the above way completely discrete quantum measure \cite{Kha4} can be
written as a result for vacuum expectations of the functions of the field variables
$v$, $\Omega$. Upon passing to integration over imaginary areas with the help of the
formal replacement of the tensors of a certain subset of areas $\pi$ over which
integration in the path integral is to be performed, $$\pi \rightarrow -i\pi,$$ the
result reads
\begin{eqnarray}                                                                    
\label{VEV2}%
<\Psi (\{\pi\},\{\Omega\})> & = & \int{\Psi (-i\{\pi\}, \{\Omega\})\exp{\left (-\!
\sum_{\stackrel{t-{\rm like}}{\sigma^2}}{\tau _{\sigma^2}\circ
R_{\sigma^2}(\Omega)}\right )}}\nonumber\\
 & & \hspace{-20mm} \cdot \exp{\left (i
\!\sum_{\stackrel{\stackrel{\rm not}{t-{\rm like}}}{\sigma^2}} {\pi_{\sigma^2}\circ
R_{\sigma^2}(\Omega)}\right )}\prod_{\stackrel{\stackrel{\rm
 not}{t-{\rm like}}}{\sigma^2}}{\rm d}^6
\pi_{\sigma^2}\prod_{\sigma^3}{{\cal D}\Omega_{\sigma^3}} \nonumber\\ & \equiv &
\int{\Psi (-i\{\pi\},\{\Omega\}){\rm d} \mu_{\rm area}(-i\{\pi\},\{\Omega\})},
\end{eqnarray}

\noindent where ${\cal D}\Omega_{\sigma^3}$ is the Haar measure on the group SO(4) of
connection matrices $\Omega_{\sigma^3}$. Appearance of some set ${\cal F}$ of
triangles $\sigma^2$ integration over area tensors of which is omitted (denoted as
"$t$-like" in (\ref{VEV2}))is connected with that integration over {\it all} area
tensors is generally infinite, in particular, when normalizing measure (finding
$<1>$). Indeed, different $R_{\sigma^2}$ for $\sigma^2$ meeting at a given link
$\sigma^1$ are connected by Bianchi identities \cite{Regge}. Therefore for the
spacetime of Minkowsky signature (when exponent is oscillating over all the area
tensors) the product of $\delta^6(R_{\sigma^2} - \R_{\sigma^2})$ for all these
$\sigma^2$ which follow upon integration over area tensors for these $\sigma^2$
contains singularity of the type of $\delta$-function squared. To avoid this
singularity we should confine ourselves by only integration over area tensors on those
$\sigma^2$ on which $R_{\sigma^2}$ are independent, and complement ${\cal F}$ to this
set of $\sigma^2$ are those $\sigma^2$ on which $R_{\sigma^2}$ are by means of the
Bianchi identities functions of these independent $R_{\sigma^2}$. Let us adopt regular
way of constructing 4D Regge structure of the 3D Regge geometries (leaves) of the same
structure. The $t$-{\it like} edges connect corresponding vertices in the neighboring
leaves (do not mix with the term "timelike" which is reserved for the local frame
components). The {\it diagonal} edges connect a vertex with the neighbors of the
corresponding vertex in the neighboring leaf. The $t$-{\it like} simplices (in
particular, $t$-like triangles) are then defined as those containing $t$-like edges;
the {\it leaf} simplices are those completely contained in the leaf; the {\it
diagonal} simplices are all others. It can be seen that the set of the $t$-like
triangles is fit for the role of the above set ${\cal F}$. In the case of general 4D
Regge structure we can deduce that the set ${\cal F}$ of the triangles with the
Bianchi-dependent curvatures pick out some one-dimensional field of links, and we can
simply take it as definition of the coordinate $t$ direction so that ${\cal F}$ be
just the set of the $t$-like triangles. Also existence of the set ${\cal F}$ naturally
fits our initial requirement that limiting form of the full discrete measure when any
one of the coordinates (not necessarily $t$!) is made continuous by flattening the
4-simplices in the corresponding direction should coincide with Hamiltonian path
integral (with that coordinate playing the role of time). Namely, in the Hamiltonian
formalism absence of integration over area tensors of triangles which pick out some
coordinate $t$ ($t$-like ones) corresponds to some gauge fixing.

Given the above (spontaneously arisen) asymmetry between the different area tensors we
nevertheless can ask about maximally symmetrical form of the measure extended by
inserting possible integrations \cite{Kha7}. We can integrate over ${\rm d}^6
\tau_{\sigma^2}$ in a nonsingular way if $\delta$-functions are inserted which fix the
scale of these tensors, say, at the level $\varepsilon$ $\ll$ 1. If the number of
these $\delta$-functions is 4 per vertex, on physical hypersurface of ordinary RC this
corresponds to fixing 4 degrees of freedom connected with lapse-shift vectors. The
latter define location of the next in $t$ 3D leaf relative to the current leaf. Their
continuum version in the Arnowitt-Deser-Misner Hamiltonian approach in the continuum
GR are nondynamical variables and can be chosen by hand.
Further, area tensor of $\sigma^2$ could be defined in any one of the 4-simplices
$\sigma^4$ $\supset$ $\sigma^2$, and the more detailed notation is
$v_{\sigma^2|\sigma^4}$. Above the $v_{\sigma^2}$ means $v_{\sigma^2|\sigma^4}$ at
some $\sigma^4$ = $\sigma^4(\sigma^2)$ $\supset$ $\sigma^2$ (function of $\sigma^2$).
Insert ${\rm d}^6v_{\sigma^2|\sigma^4}$ for {\it all} $\sigma^4$ $\supset$ $\sigma^2$.
As applied to functions of the above $v_{\sigma^2}$ in the frames of certain
$\sigma^4(\sigma^2)$ only, the new integrations over ${\rm d}^6v_{\sigma^2|\sigma^4}$,
$\sigma^4$ $\neq$ $\sigma^4(\sigma^2)$, simply contribute into a normalization factor
(some intermediate regularization is implied which sets large but finite values for
the integration limits over area tensors). Such the extended form of the measure is
just used in the following when passing to physical hypersurface (of the ordinary RC).

There is the invariant (Haar) measure ${\cal D}\Omega$ in (\ref{VEV2}) which looks
natural from symmetry considerations. From the formal point of view, in the
Hamiltonian formalism (when one of the coordinates is made continuous) this arises
when we write out standard Hamiltonian path integral for the Lagrangian with the
kinetic term $\pi_{\sigma^2}\circ\Om_{\sigma^2}\dot{\Omega}_{\sigma^2}$
\cite{Kha3,Kha4}. To this end, one might pass to the variables $\Omega_{\sigma^2}
\pi_{\sigma^2}\!$ = $\!P_{\sigma^2}$ and $\Omega_{\sigma^2}$ (in 3D case used in
\cite{Wael,Kha3}). The kinetic term $P\dot{\Omega}$ with arbitrary matrices $P$,
$\Omega$ leads to the standard measure ${\rm d}^{16}P{\rm d}^{16}\Omega$, but there
are also $\delta$-functions taking into account II class constraints to which $P$,
$\Omega$ are subject, $\delta^{10}(\Om\Omega - 1)\!$ $\!\delta^{10}(\Om P +
\bar{P}\Omega)$. Integrating out these just gives ${\rm d}^6\pi{\cal D}^6\Omega$.
Following our strategy of recovering full discrete measure from requirement that it
reduces to the Hamiltonian path integral whatever coordinate is made continuous, the
same Haar measure should be present also in the full discrete measure.

One else specific feature of the quantum measure is the absence of the inverse
trigonometric function 'arcsin' in the exponential, whereas the Regge action
($\ref{S-RegCon}$) contains such functions. This is connected with using the canonical
quantization at the intermediate stage of derivation: in gravity this quantization is
completely defined by the constraints, the latter being equivalent to those ones
without $\arcsin$ (in some sense on-shell).

The theory with independent area tensors is locally trivial (just as 3D RC). In the
considered formalism this explicitly exhibits at the negligibly small values of
$\tau_{\sigma^2}$ when we get factorisation of the quantum measure obtained into the
"elementary" measures on separate areas of the type
\begin{equation}                                                                    
\label{separate} \exp{(i\pi\circ R)}{\rm d}^6\pi{\cal D}R.
\end{equation}

\noindent Upon splitting antisymmetric matrices ($\pi$ and generator of $R$) into
self- and antiselfdual parts like
\begin{eqnarray}                                                                    
\pi_{ab} & \equiv & {1\over 2}\ppi_k\pSig^k_{ab} + {1\over 2}\mpi_k\mSig^k_{ab} \\
\pmR & = & \exp (\pmbphi\pmbSig) = \cos \pmphi + \pmbSig\pmbn\sin \pmphi \nonumber
\end{eqnarray}

\noindent ($\pmbn$ = $\pmbphi / \pmphi$ is unit vector and the basis of self- and
antiselfdual matrices $i\pmSig^k_{ab}$ obeys the Pauli matrix algebra) the measure
(\ref{separate}) splits as
\begin{equation}                                                                   
\label{+d-mu-d-mu}%
\exp(i\,^+\!\pi\circ\,^+\! R)\d^3\pbpi{\cal D}\,^+\! R\cdot \exp(i\,^-\!\pi\circ\,^-\!
R)\d^3\mbpi{\cal D}\,^-\! R
\end{equation}

\noindent where ${\cal D}\pmR$ = $\left (4\pi^2\pmphi^2\right )^{-1}\sin^2\pmphi {\rm
d}^3\pmbphi$. When calculating expectations of powers of area vectors $\pmbpi$
integrals over ${\rm d}^3\pmbpi$ give (derivatives of) $\delta$-functions which are
then easily integrated out giving
\begin{eqnarray}                                                                   
\label{<bpi>}%
<(\pmbpi)^{2k}> & = & const\cdot \int (-i\pmbpi)^{2k} {\rm d}^3\pmbpi \int
e^{\textstyle i\pmbpi \circ \pmR} {\cal D}\pmR \nonumber\\ & = & const\cdot \int \left
[ \partial^{2k}_{\pmbr} \delta^3(\pmbr ) \right ] {{\rm d}^3\pmbr \over
\sqrt{1-\pmbr^2}} = {4^{-k}(2k+1)!(2k)!\over k!^2}
\end{eqnarray}

\noindent where $\pmbr$ = $\pmbn \sin \pmphi$, and "const" is a normalization factor.
Knowing how monomials are averaged we can select the measure needed for that; thereby
the result extends to averaging arbitrary polynomial or, by continuity, practically
arbitrary function,
\begin{eqnarray}                                                                   
\label{Euclidean_measure}%
<f(\pi)> & = & \int{f (-i\pi ){\rm d}^6\pi\int{e^{\textstyle i\pi\circ R}{\cal D}R}}
\nonumber\\ & = & \int{f (\pi ) {\nu (|\pbpi |)\over |\pbpi |^2}{\nu (|\mbpi |)\over
|\mbpi |^2} {\d^3\pbpi\over 4\pi} {\d^3\mbpi\over 4\pi}}, \\ & & \nu(s)={s\over\pi}
\int\limits_{-\infty}^{+\infty}\exp{(-s\cosh \eta)}\cosh \eta \, {\rm d} \eta =
{2s\over \pi }K_1(s). \nonumber
\end{eqnarray}

\noindent $K_1$ is the modified Bessel function. (A shorter way to get the same is to
proceed by moving integration contours over curvatures to complex plane \cite{Kha5}.)

Next, as considered below the equation (\ref{v*v}), we are aiming at implementing
kinematical relations of that type in order to get quantum measure in the genuine
ordinary RC from the obtained measure in area tensor RC. For that we find the measure
of interest as the result of reducing the measure obtained in the superspace of
independent area tensors onto the hypersurface $\Gamma_{\rm Regge}$ corresponding to
the ordinary RC in this superspace. The quantum measure can be considered as a linear
functional $\mu_{\rm area}(\Psi)$ on the space of functionals $\Psi (\{v\})$ on the
configuration space (for our purposes here it is sufficient to restrict ourselves to
the functional dependence on the area tensors $\{v\}$; the dependence on the
connections is unimportant). The physical assumption is that we can consider ordinary
RC as a kind of the state of the more general system with independent area tensors.
This state is described by the following functional,
\begin{equation}                                                                   
\Psi (\{v\}) = \psi (\{v\})\delta_{\rm Regge}(\{v\}),
\end{equation}

\noindent where $\delta_{\rm Regge}(\{v\})$ is the (many-dimensional)
$\delta$-function with support on $\Gamma_{\rm Regge}$. The derivatives of
$\delta_{\rm Regge}$ have the same support, but these violate positivity in our
subsequent construction. To be more precise, delta-function is distribution, not
function, but can be treated as function if regularised. If the measure on such the
functionals exists in the limit when regularisation is removed, this allows to define
the quantum measure on $\Gamma_{\rm Regge}$,
\begin{equation}                                                                   
\label{mu-proj}%
\mu_{\rm Regge}(\cdot) = \mu_{\rm area}(\delta _{\rm Regge}(\{v\})~\cdot).
\end{equation}

Uniqueness of the construction of $\delta_{\rm Regge}$ follows under quite natural
assumption of the minimum of lattice artefacts. Let the system be described by the
metric $g_{\lambda\mu}$ constant in each of the two 4-simplices $\sigma^4_1$,
$\sigma^4_2$ separated by the 3-face $\sigma^3$ = $\sigma^4_1\cap \sigma^4_2$ formed
by three 4-vectors $\iota^\lambda_a$. These vectors also define the metric induced on
the 3-face, $g^{\|}_{ab}$ = $\iota^\lambda_a\iota^\mu_bg_{\lambda\mu}$. The continuity
condition for the induced metric is taken into account by the $\delta$-function of the
induced metric variation,
\begin{equation}                                                                   
\label{delta-g} \Delta_{\sigma^3}g^{\|}_{ab}\stackrel{\rm def}{=}
g^{\|}_{ab}(\sigma^4_1)-g^{\|}_{ab}(\sigma^4_2).
\end{equation}

\noindent As for the $\delta_{\rm Regge}$, it is of course defined up to a factor
which is arbitrary function nonvanishing at nondegenerate field configurations. In the
spirit of just mentioned principle of minimizing the lattice artefacts it is natural
to choose this factor in such the way that the resulting $\delta$-function factor
would depend only on hyperplane defined by the 3-face but not on the form of this
face, that is, would be invariant with respect to arbitrary nondegenerate
transformations $\iota^\lambda_a$ $\mapsto$ $m^b_a\iota^\lambda_b$. To ensure this,
the $\delta$-function should be multiplied by the determinant of $g^{\|}_{ab}$
squared. This gives
\begin{equation}                                                                   
\label{inv-delta}%
[{\rm det}(\iota^\lambda_a\iota^\mu_bg_{\lambda\mu})]^2
\delta^6(\iota^\lambda_a\iota^\mu_b \Delta_{\sigma^3}g_{\lambda\mu}) = V^4_{\sigma^3}
\delta^6(\Delta_{\sigma^3}S_{\sigma^3}).
\end{equation}

\noindent Here $S_{\sigma^3}$ is the set of the 6 edge lengths squared of the 3-face
$\sigma^3$, $V_{\sigma^3}$ is the volume of the face.

Further, the product of the factors (\ref{inv-delta}) over all the 3-faces should be
taken. As a result, we have for each edge the products of the $\delta$-functions of
the discontinuity of its length between the 4-simplices taken along closed contours,
$\delta (s_1-s_2)\delta (s_2-s_3)\dots\delta (s_N-s_1)$ containing singularity of the
type of the $\delta$-function squared. In other words, the conditions equating
(\ref{delta-g}) to zero on the different 3-faces are not independent. The more
detailed consideration allows us to cancel this singularity in a way symmetrical with
respect to the different 4-simplices (thus extracting irreducible conditions), the
resulting $\delta$-function factor remaining invariant with respect to arbitrary
deformations of the faces of different dimensions keeping each face in the fixed plane
spanned by it \cite{Kha6}.

Besides factors (\ref{inv-delta}), we need to impose the conditions ensuring that
tensors of the 2-faces in the given 4-simplex define a metric in this simplex
\cite{Kha7}. These conditions of the type of (\ref{v*v}) can be easily written in
general form. Let a vertex of the given 4-simplex be the coordinate origin and the
edges emitted from it be the coordinate lines $\lambda$, $\mu$, $\nu$, $\rho$, \dots =
1, 2, 3, 4. Then the (ordered) pair $\lambda\mu$ means the (oriented) triangle formed
by the edges $\lambda$, $\mu$. The conditions of interest take the form
\begin{equation}                                                                   
\label{tetrad} \epsilon_{abcd}v^{ab}_{\lambda\mu}v^{cd}_{\nu\rho} - {1 \over 4!} \left
(\epsilon^{\xi\sigma\varphi\chi}\epsilon_{abcd}v^{ab}_{\xi\sigma}v^{cd}_{\varphi\chi}\right
) \epsilon_{\lambda\mu\nu\rho} = 0.
\end{equation}

\noindent This expresses proportionality of
$\epsilon_{abcd}v^{ab}_{\lambda\mu}v^{cd}_{\nu\rho}$ to
$\epsilon_{\lambda\mu\nu\rho}$. The LHS is symmetric 6 $\!\times\!$ 6 matrix w.r.t.
the antisymmetric pairs $\lambda\mu$, $\nu\rho$. It has 21 different nontrivial
elements of which 20 are independent ones (contraction with
$\epsilon^{\lambda\mu\nu\rho}$ gives identical zero). These 20 equations define the
16-dimensional surface $\gamma (\sigma^4)$ in the 36-dimensional configura\-tion space
of the six antisymmetric tensors\footnote{There are also the linear constraints of the
type $\sum{\pm v}$ = 0 providing closing surfaces of the 3-faces of our 4-simplex.
These are assumed to be already resolved.\label{sum-v}} $v^{ab} _{\lambda\mu}$. The
factor of interest in quantum measure is the product of the $\delta$-functions with
support on $\gamma (\sigma^4)$ over all the 4-simplices $\sigma^4$. The covariant form
of the constraints (\ref{tetrad}) with respect to the world index means that the
product of these $\delta$-functions in each the 4-simplex is the scalar density of a
certain weight with respect to the world index, that is, the scalar up to power of the
4-volume $V_{\sigma^4}$. Therefore introducing the factors of the type
$V_{\sigma^4}^{\eta}$ we get the scalar at some parameter $\eta$. Namely, the product
of the factors
\begin{equation}                                                                   
\label{delta-metric}%
\prod_{\sigma^4}{\int{V_{\sigma^4}
^{\eta}\delta^{21}(\epsilon_{abcd}v^{ab}_{\lambda\mu | \sigma^4}v^{cd}_{\nu\rho
|\sigma^4} - V_{\sigma^4}\epsilon _{\lambda\mu\nu\rho})\,{\rm d}V_{\sigma^4}}}
\end{equation}

\noindent at $\eta$ = 20 is the world index scalar, i.e. invariant w.r.t. arbitrary
deformations of the 4-simplex $v^{ab}_{\lambda\mu}$ $\mapsto$
$\xi^\nu_\lambda\xi^\rho_\mu v^{ab}_{\nu\rho}$ as is desirable from the viewpoint of
minimization of the lattice artefacts. The (\ref{delta-metric}) is a short symmetrical
way to write our 20 irreducible conditions. Here $V_{\sigma^4}$ serves as dummy
variable, integrating over it simply excludes $V_{\sigma^4}$ from 21 conditions and
yields 20 delta functions of 20 independent conditions per $\sigma^4$.

Qualitatively, it is important that $\delta$-factors (\ref{inv-delta}),
(\ref{delta-metric}) automatically turn out to be invariant w.r.t. the overall
rescaling area tensors. Therefore introducing these into the measure functional
(\ref{mu-proj}) turns out to keep convergence properties of the corresponding
integrals. The integrals convergent in area tensor RC remains convergent on physical
hypersurface of ordinary RC (both at infinite or at infinitely small area tensors). To
be exact, invariance of the additional factors w.r.t. the scaling only
$\pi_{\sigma^2}$ is needed for that. As these stand, these factors possess this
property. For example, it is seen for (\ref{inv-delta}) upon rewriting it in terms of
the triples of area vectors of the 3-face $\sigma^3$ = $\sigma^4_1 \cap \sigma^4_2$,
namely, $\bv^{(1)}_1$, $\bv^{(1)}_2$, $\bv^{(1)}_3$ defined in $\sigma^4_1$ and
$\bv^{(2)}_1$, $\bv^{(2)}_2$, $\bv^{(2)}_3$ defined in $\sigma^4_2$, as $\left
[\bv^{(1)}_1 \times \bv^{(1)}_2 \cdot \bv^{(1)}_3\right ]^4 \delta^6 \left
(\bv^{(1)}_{\alpha} \cdot \bv^{(1)}_{\beta} - \bv^{(2)}_{\alpha} \cdot
\bv^{(2)}_{\beta} \right )$ (modulo (\ref{delta-metric}) there is no matter whether
$\bv$ means $\pbv$ or $\mbv$ here). Violation of this property of invariance w.r.t.
rescaling $\pi_{\sigma^2}$ might arise when some $\pi_{\sigma^2}$ in these factors are
expressed (see footnote on page \pageref{sum-v}) as an algebraic sum of some another
$\pi_{\sigma^2}$ chosen as independent variables plus some $\tau_{\sigma^2}$. However,
the role of this circumstance is that different $\pi_{\sigma^2}$ on physical
hypersurface cannot achieve 0 simultaneously, and convergence properties of the
measure cannot become worse when passing from area tensor to ordinary RC.

Now consider to what extent our system can be similar to the ordinary lattice field
theory in which correlators (Green functions) are well defined due to the lattice
regularization, UV cut off being determined by the (fixed) lattice spacing. Knowing
the lengths expectation values is not sufficient to make conclusion on possible
finiteness of the theory, we need a more detailed study of the probability
distribution for the linklengths, that is, of the quantum measure.

Indeed, let the linklengths are allowed to be arbitrarily close to zero with some
probability. (This does not contradict to the statement on their expectation values
being finite and nonzero.) We may speak of the {\it dynamical} lattice with spacings
being dynamical variables (linklengths) themselves. In fact, we have an ensemble of
the lattices with different spacings. If linklengths can be found with noticeable
probability in the arbitrarily small neighborhood of zero, this means that the
ensemble includes the lattices in the limit of zero spacings, that is, in the limit of
the regularization removed, and finiteness of the theory is not evident.

On the contrary, let the quantum measure has the support strictly separated from zero
lengths. In this case the theory is thought to be finite like a lattice theory with
the difference that now the lattice is the {\it dynamical} one. (At large
areas/lengths suitable properties are provided by the exponential cut off in the
measure.)

Qualitative considerations lead to namely this last possibility. Let us make a simple
scaling estimate. Consider estimation model used in \cite{Kha7}. The half of 36
components of the 6 antisymmetric tensors $v^{ab}_{\lambda\mu}$ in a given 4-simplex
are dynamical $\pi^{ab}_{\lambda\mu}$ ($\lambda\mu$ = 12, 23, 31), another half are
$\tau^{ab}_{\lambda\mu}$ ($\lambda\mu$ = 14, 24, 34, and a scale $\varepsilon\!$
$\!\ll\!$ 1 for $\tau_{\sigma^2}$ is chosen). Denote by $x$ a scale of tensors
$\pi_{\sigma^2}$ in the given 4-simplex. Then ${\rm d}^{18}\pi^{ab}_{\lambda\mu}$
behaves like $x^{17}{\rm d}x$ (also together with the factor (\ref{delta-metric}) due
to its invariance w.r.t. the rescaling $\pi_{\sigma^2}$). Besides that, there is the
factor in the measure (\ref{Euclidean_measure}) which gives $x^{-4}e^{-x}$ for each
leaf/diagonal triangle (one might write $e^{-\lambda x}$ with $\lambda$ = $O(1)$ but
$x$ is itself defined up to a value of the order of unity). Finally, the factors like
(\ref{inv-delta}) serve to equate the scales $x_1$ and $x_2$ in the 4-simplices
sharing a 3-face and in our scaling estimate effect of such factor is equivalent to
the effect of $x_1\delta (x_1 - x_2)$ on the two measures $f_1(x_1){\rm d}x_1$ and
$f_2(x_2){\rm d}x_2$: $f_1(x_1){\rm d}x_1f_2(x_2){\rm d}x_2x_1\delta (x_1 - x_2)$
$\Rightarrow$ $xf_1(x)f_2(x){\rm d}x$. Collecting together just considered factors in
the measure according to this rule we find
\begin{equation}\label{dx}                                                         
(e^{-x}x^{-4})^{L_2}x^{18N_4}{\rm d}x/x = (e^{-x}x^8)^{3N_4/2}{\rm d}x/x
\end{equation}

\noindent where $L_2$ is the number of the leaf/diagonal triangles and $N_4$ is the
number of the 4-simplices. We have used simple combinatorial relation $N_4/L_2$ =
$2/3$ \cite{Kha7} for the above considered (after eq. (\ref{VEV2})) regular way of
constructing 4D Regge structure from the 3D ones. As $N_4$ grows, the factor in
(\ref{dx}) approaches $\delta (x - 8)$.

Thus, the presumable picture looks as if the main contribution to the path integral
were from linklengths being the sum of some constant (uniform) part $l_0$ of the order
of Plank scale plus small fluctuations around it. The finite nonzero $l_0$ provides
finite calculational framework for a correlator (say, of two or more defect angles
or bivectors located at a certain distances), as if it were considered on the lattice
with fixed spacing $l_0$.

Of course, the more detailed analytical and, probably, numerical investigations are in
order to confirm this picture.

I am grateful to I.B. Khriplovich for attention to the work and discussion. The
present work was supported in part by the Russian Foundation for Basic Research
through Grant No. 05-02-16627-a.

\end{document}